\documentclass[
    ,final            
  ]{aipproc}

\layoutstyle{8x11single}

\begin{document}

\title{Understanding low energy reaction with exotic nuclei}

\author{F.M. Nunes, N.C. Summers}{
  address={NSCL and Department of Physics and Astronomy,  
  MSU, East Lansing MI 48824,  USA}
}

\author{A.M. Moro}{
  address={Departamento de FAMN, Universidad de Sevilla, Aptdo. 
1065, 41080 Sevilla, Spain}
}

\author{A.M. Mukhamedzhanov}{
  address={Cyclotron Institute, Texas A\& M University, College Station TX
77843 USA}
}

\begin{abstract}
Recent developments on the understanding of low energy reactions are
highlighted. Emphasis is given to the CDCC framework where the
breakup channels of the projectile are included explicitly.
Properties of the breakup couplings are presented. 
Comments are given with regard to the separation between
the nuclear and the Coulomb contributions to breakup cross sections
as well as the dependence on the optical potentials. A discussion
on the sensitivity of the CDCC basis is discussed,
by comparing pure breakup results with transfer to 
the continuum calculations. Finally, some remaining controversies 
show the need to go beyond the single particle picture for
the projectile.
\end{abstract}

\maketitle


\section{Introduction}

Light nuclei on the driplines can be studied
through a variety of reactions. Models
for nuclear reactions have been developed in recent
years to incorporate the exotic features
of these dripline nuclei \cite{rev}. They include the effects of 
the long tails of the wavefunctions, the correct asymptotics, and 
the proximity of the ground state to threshold. Whereas in the high energy
regime many approximations are appropriate, the real challenge for
reaction theory lies in the low energy regime
where most approximations are not valid. 

It is in the low
energy region (5-50 MeV/A) that observables become much more
sensitive to the detailed structure of the projectile and
where more can be learnt. It is also at low energy where
there is a larger sensitivity to the details of the
interaction with the target and where more care needs
to be taken in modelling the reaction. 

We consider dripline nuclei of two body nature, meaning
that the projectile can be decomposed into a core and
a valence nucleon. Then, the study of the reaction consists of a
three body scattering problem. Due to the loosely bound nature of
the projectile, three body effects  need to be
carefully considered in the lower energy regime.
The exact way to formulate this problem would be to use 
Integral Faddeev Equations. However, due to technical problems 
the Continuum Discretized Coupled Channel Method (CDCC) \cite{cdcc} 
is the best working alternative. In CDCC, the continuum couplings 
are included to all orders and nuclear and Coulomb are treated
consistently.
 
The work here presented is based on the CDCC framework.
We first present the properties of the couplings in breakup reactions, 
in particular continuum-continuum couplings. Second we emphasize
the difficulty in separating nuclear from Coulomb contributions
and finally we discuss the choice of the Jacobi coordinates to
represent the CDCC basis. Finally we make some comments on
lingering controversies calling for better description of
the projectile.

\section{Couplings in the continuum}

The proximity to the breakup threshold has been shown to have 
important effects
in the reaction mechanism \cite{nunes99}. 
Continuum couplings are a way of
looking into the effect of the final state
interactions, an integral part of CDCC. They consist of the
sum of the  core-target interaction with the fragment-target
interaction (both Coulomb and nuclear), averaged over an
initial and a final bin wavefunction. A bin wavefunction
is essentially a scattering wavefunction describing the
two body continuum of the projectile, but averaged over a
finite energy segment \cite{nunes99}. 

\begin{figure}
  \includegraphics[height=.34\textheight]{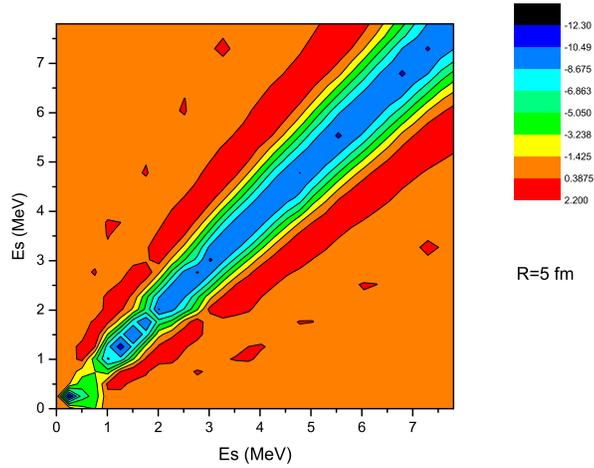}
  \caption{Contour plot of the continuum continuum potential couplings of an
  l=0 transition between two s-waves as a function of the initial and 
  final energies.(color)}
  \label{contss}
\end{figure}
\begin{figure}
  \includegraphics[height=.34\textheight]{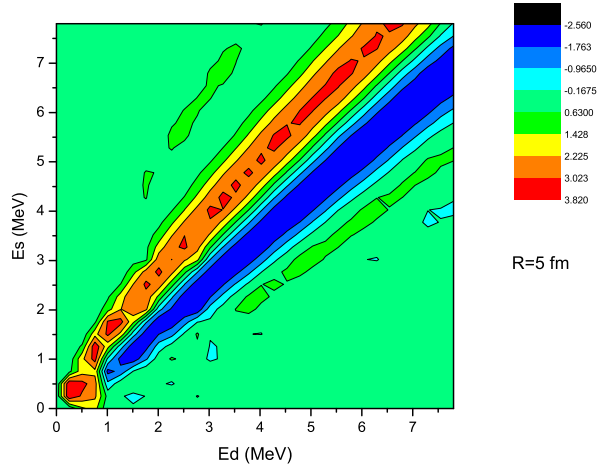}
  \caption{Contour plot of the continuum continuum potential couplings of an
  l=2 transition between an initial s-waves with energy $E_i$ and
  a final d-wave with energy $E_f$.(color)}
  \label{contsd}
\end{figure}
The properties of these continuum couplings
and the influence they can have on breakup observables were
addressed in detail \cite{nunes04}.  The couplings considered
are those involved in the breakup of $^8$B into $^7$Be+p
when impinging on a $^{58}$Ni target at 25.8 MeV \cite{nunes99}.
Continuum couplings are most important when the initial and
final state energies ($E_i$ and $E_f$) are close to each other. 
The energy here referred to are the relative energy of the
projectile in the breakup state $^7$Be+p. If the initial
and final states have the same centrifugal barrier and the transition
is a monopole transition, then the
couplings are only non-zero when the initial and final energies
match. 
This corresponds to the condition of orthonormality between
bin states. The orthogonality condition is illustrated
very clearly in Fig. \ref{contss} where a contour plot of the coupling 
potentials is shown for initial and final s-wave bins.
As the difference in the centrifugal barrier  and
the order of the transition increases, one obtains
a wider region $E_i- \Delta <E_f < E_i+\Delta$ where
contributions are significant. In Fig. \ref{contsd} we show a
contour plot for an initial d-wave with energy $E_d$ to a 
final s-wave with energy $E_s$ (reverse
coupling are equivalent).
It is clear the formation of ridges parallel to the $E_i=E_f$ line.
These couplings are attractive for $E_d < E_s$ and repulsive
for $E_d > E_s$.

The fact that continuum couplings are only relevant around
a certain region of the energy space offers a hint to
optimize the calculations. We have performed tests
for a couple of examples, namely the above mentioned example, 
the $^8$B on Ni experiment from Notre Dame \cite{nd}, 
and another case to be discussed in the next section, 
the $^7$Be breakup on Pb  experiment from Michigan State University 
\cite{summ04}. In both cases we found
up to 30\% time gain for the calculations where only the lower l=0,1,2 
projectile partial waves were included. For the larger calculations
where l=3 is also included (necessary for complete convergence) 
the order of some transitions become too large to justify
a truncation along the $E_i=E_f$ line. Unfortunately,
in our typical calculations, it is exactly the larger partial
waves (l>2) that make the calculations very large and
increase running time dramatically. It is expected that
this optimization will be more helpful when including
core degrees of freedom as then even for the lower
partial waves due to an order of magnitude increase
in the number of channels \cite{summ05}.

\section{Separation between nuclear and Coulomb}

Historically, there has always been the underlying assumption
that, by appropriately choosing the experimental conditions,
Coulomb effects can be isolated from nuclear effects.
Especially when breakup reactions are used to extract Astrophysical
information, such as radiative capture
rates, this separation is crucial \cite{davids98}. In \cite{summ04}
breakup of $^7$Be on a heavy and a light target is
considered, motivated by recent experimental plans.
The breakup reaction on Pb would be Coulomb dominated and would
allow to extract information on the astrophysical factor $S_{34}$,
whereas the experiment on the carbon target would be
driven by nuclear effects and would provide an asymptotic normalization
coefficient for the $\alpha+^3$He system, again linking back
to the astrophysical capture reaction at zero energy.

Results of CDCC calculations from \cite{summ04}, include
the continuum of $^7$Be to all orders. 
The most important conclusion
of that work is that a simple angular selection of the so-called 
Coulomb Dissociation is not sufficient to guarantee that
the data is nuclear free. Identically, for the lighter target,
the data is always contaminated by a Coulomb contribution.
Also, for the carbon case, coupling effects were in general
non negligible for the center of mass forward angular region.
The work in \cite{summ04} shows that only careful massaging of the 
data, i.e. specific three-body kinematic selections, may recover the
purity that is desired for Astrophysical problems.

\begin{figure}[t]
\resizebox*{0.3\textheight}{!}{\includegraphics{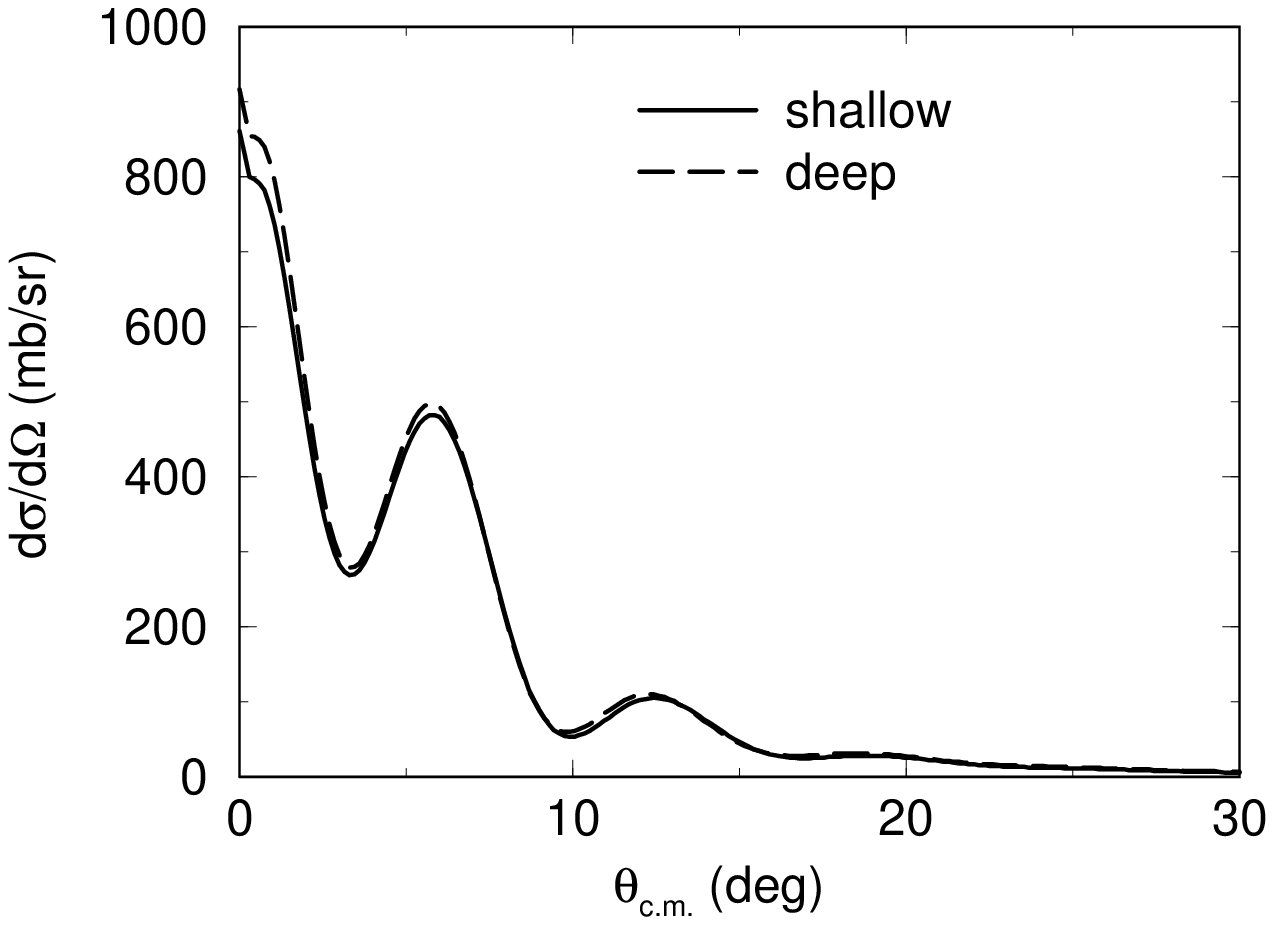}}
\hspace{1.cm}
\resizebox*{0.3\textheight}{!}{\includegraphics{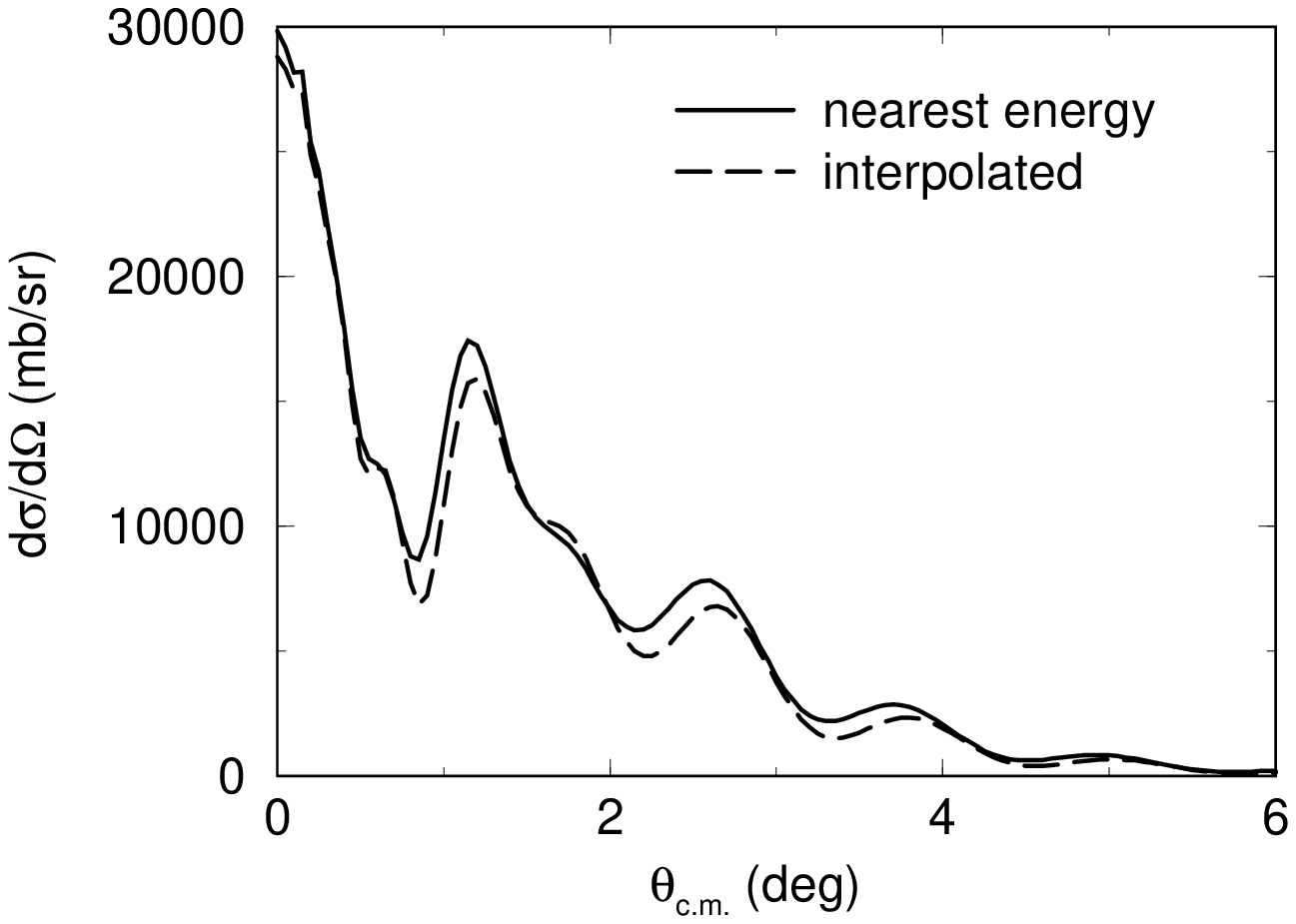}}
\vspace{-0.4cm}
\caption{\label{opt} Sensitivity to optical potentials
for the breakup of $^7$Be on C (left) and Pb (right).}
\end{figure}

It is common understanding that optical potentials can produce 
large uncertainties and  there is a preoccupation in either
keeping nuclear contributions small or choosing reaction
regimes where there is less sensitivity to the details of
the optical potentials. From the two cases studied in \cite{summ04}
one expects that the carbon case
will show a larger dependence given that it has a larger
nuclear component when compared to it Coulomb component. 
For illustration purposes we concentrate on
the $\alpha$-target interaction. 
We have compared the results when using
a shallow \cite{hauser} or a deep potential fit \cite{smith} 
for the $\alpha-^{12}$C
and find a minor effect (see Fig.\ref{opt} left).
For the heavier target there is a weak dependence on the
optical potential and the only issue arising has to do with the fact
that optical potentials are not available at the correct
energy. We show the sensitivity to the energy choice in 
Fig. \ref{opt}. The differential cross section is plotted for
the case where the $\alpha$-$^{208}$Pb potential
is taken directly from the literature \cite{boonin} at the nearest 
available energy and  compare to the results when an interpolation of the
potentials is made to the correct energy.
As can be seen the dependence is very small.
These results suggest that, when the scattering of the fragments 
is well understood, 
the optical potentials themselves do not
introduce significant ambiguities in the analysis.

\section{Comparing CDCC bases}

\begin{figure}[t]
  \includegraphics[height=.15\textheight]{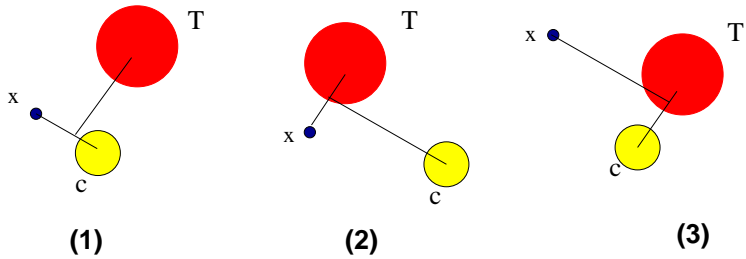}
  \caption{Each Faddeev component is written in
  its corresponding Jacobi coordinate system.}
  \label{jacobi}
\end{figure}

A variety of breakup models are presently in use and, when two different 
models are applied to the same problem, there is
often a disparity in the predictions. 
In this sense, a generalized effort to bridge the various approaches 
is very much needed. One of the important issues lies in the choice
of the coordinate representation of the continuum wavefunctions.
As mentioned in the introduction, the problem of a two-body projectile
impinging on a target consists of a three-body problem of
which an exact solution would be obtained by solving the
Integral Faddeev Equations. Then, the wavefunction would contain
components in the three Jacobi coordinates represented in Fig. \ref{jacobi}. 
Due to the complexity of this task,
the CDCC method was derived \cite{austern}. However, the CDCC
method already imposes a preferential representation of
the continuum, namely that of the projectile (coordinate set
(1) in Fig. \ref{jacobi}).
If there are important resonant
states in the target-fragment subsystem, 
the Jacobi coordinate set (2) in Fig. \ref{jacobi} would become
more appropriate and the representation in terms of coordinates (1)
would probably be very difficult.

\begin{figure}[b]
\resizebox*{0.15\textheight}{!}{\includegraphics{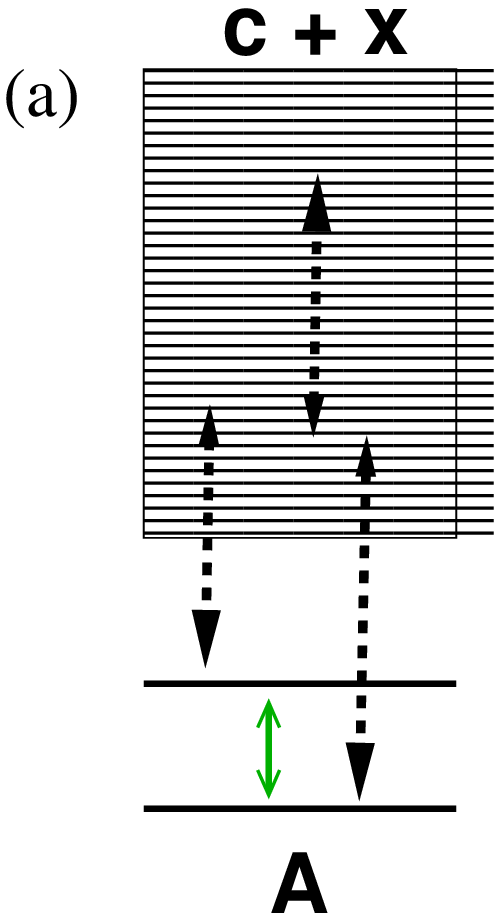}}
\hspace{1.cm}
\resizebox*{0.28\textheight}{!}{\includegraphics{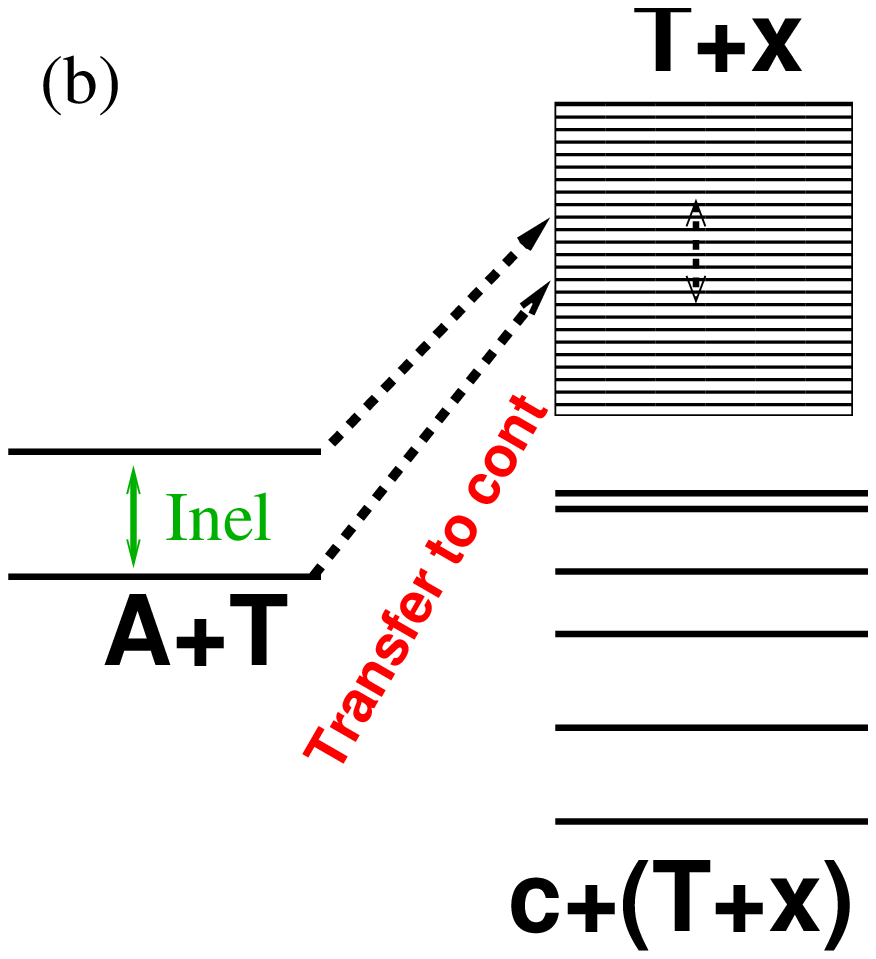}}
\vspace{-0.4cm}
\caption{\label{coup} Breakup couplings for a two body projectile 
(c+x) impinging on a target T (on the left) and corresponding
transfer to the continuum couplings (on the right).}
\end{figure}

The standard CDCC breakup uses coordinates (1) and
the couplings are single particle excitations of the
projectile into the continuum (referred to as BU). These are illustrated in Fig. \ref{coup} left.
where the projectile A is excited into A$^*=c+x$ through the interaction
with the target T. 
Alternatively, one can imagine that the projectile transfers
its valence particle x into the continuum of the target
(referred to as TR*).
CDCC would be then applicable to the continuum of the
T+x system and thus be associated with a final state interaction. 
In that situation the relevant 
coordinates would be (2) and the couplings would be
transfer couplings such as those represented in Fig. \ref{coup}(right).

\begin{figure}[t]
  \includegraphics[height=.25\textheight]{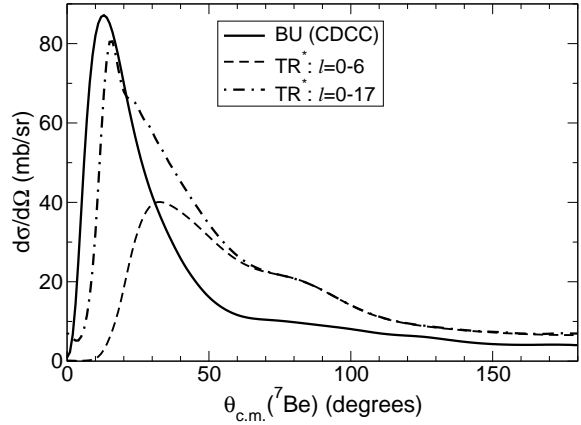}
  \caption{Angular distribution for the $^7$Be after $^8$B breakup:
  comparison between the standard breakup calculation and the
  transfer to the continuum.}
  \label{b8tr-bu}
\end{figure}

Previous work shows that both methods may hold different
results. For example, the analysis of the $^8$B breakup
\cite{nunes99} was performed within the standard CDCC
approach whereas the $^8$Li data measured in the same energy
regime  could only be explained when using transfer to the
continuum \cite{moro03}.
We have performed a comparative study between the standard
CDCC breakup approach and the so called transfer to the 
continuum \cite{moro04}. As a testing case we start with
the $^8$B breakup which is well understood within the
standard approach.
Detailed data exists for
$^8$B$\rightarrow ^7$Be$+p$ on $^{58}$Ni at 25.6 MeV \cite{nd}. 
Calculations using the standard CDCC to breakup (BU)
$^{8}$B$+^{58}$Ni $\rightarrow (^{7}$Be$+p) +^{58}$Ni
have provided very good agreement with experiment \cite{nunes99,toste01}.
One can think of the alternative path to breakup, as 
transfer to the continuum (TR*) of the $^{59}$Cu, i.e. 
$^{8}$B$+^{58}$Ni $\rightarrow ^{7}$Be$ +(p+^{58}$Ni). 
The results
for the angular distributions of $^7$Be are shown in Fig. \ref{b8tr-bu}.
BU calculations are fully converged and provide a
pronounced Coulomb peak around 10-20 degrees. This same peak is not well
reproduced with the TR* approach. In fact,
the convergence rate of the TR* calculation
is very slow and the calculations are much larger than BU, due to the
nature of the non-local transfer kernels. 
The breakup of $^8$B on $^{58}$Ni at 25.8 MeV 
is a good example where the BU configuration works
much better than the  TR* configuration.

General guidelines as to the conditions for choosing
the standard breakup approach or the transfer to the continuum
approach are under study. It seems clear that for the standard
breakup approach to be valid, the average relative energy
for c+x during the reaction should be small as well as the average
relative angular momentum for c+x. Identically if 
the transfer to the continuum is to be applicable,
the average excitation energy
for t+x should be small as well as the average
relative angular momentum for t+x. 
However the situation is not always clear.
There are also issues on the choice of certain interactions
that play a different role in the transfer process from the
breakup process. A more detailed discussion on these and other
issues will soon become available \cite{moro04}.

\section{Remaining controversies}

As breakup states are an essential path in reactions
with loosely bound projectiles, only reaction models that include
the continuum have been successful in describing measurements
for nuclei on the driplines.

There are some puzzling problems
which can well correspond to cases where a single
particle description is less appropriate.
Note that, in the reaction models we have
discussed, the ground state of the projectile is taken 
to be a single particle state of unit spectroscopic factor
produced by a simple Woods-Saxon and spin-orbit interaction
with standard geometries with a depth fitted to the correct 
binding energy of the $c+x$ system. The continuum is produced
with that same interaction to ensure orthogonality.
Below, we briefly discuss three different puzzles, the first
related to breakup experiments of $^8$B, the second related
to inelastic process with $^{11}$Be and the last associated
with knock-out measurements for $^{16}$C. They serve as
an illustration of the need to go beyond the models so far
developed.

There have been several
$^8$B breakup experiments performed at different facilities to provide the
needed information for $S_{17}$. Using our best understanding 
of the reaction mechanism, and assuming the projectile can
be represented by  $^7$Be(inert)$+p$, the Notre Dame data and 
the NSCL/MSU data show  a $60$\% inconsistency in the quadrupole excitation 
strength. This is an extremely severe problem from the point of few 
of the direct capture cross section \cite{b8neil}. 
In juxtaposition, accurate measurements have shown that $^7$Be first excited 
state contributes to the ground state
of $^8$B \cite{b8gsi}.

GANIL data of $^{10,11}$Be(p,p') inelastic scattering
have remained unpublished for the last five years
as we have been unable to understand the process \cite{lapoux}. 
Due to the proximity to the continuum, there is a large
contribution of breakup $^{10}$Be+n states to the inelastic 
cross section. Presently, this can only be modelled within an 
inert few-body model. It was not possible within this model
to understand $^{10}$Be data and the $^{11}$Be simultaneously.
Identical conclusions were found in MSU data \cite{shrivastava}.
It is well known that $^{10}$Be first excited state contributes
to the $^{11}$Be ground state \cite{winfield,aumann}.
A number of preliminary tests have been performed \cite{moro} 
and suggest that  excitation of $^{10}$Be is very important 
for $^{11}$Be(p,p') in particular in the breakup channels. 
However, within the current model, a definite conclusion
is yet to be drawn \cite{moro}.

The analysis \cite{maddalena0} of knockout  data
to extract spectroscopic factors for $^{16}$C proved to be
extremely difficult. The same reaction model that had been
so successful in a number of cases did not provide very good
agreement with the data. Efforts to check the relevance
of core excitation in  reaction models \cite{maddalena1,shyam5},
by treating it statically were also unfruitful. In those models
the core excited component is kept constant throughout the
reaction process. This approximation does not seem adequate, especially in
cases where the couplings to core excited states are strong. 

Although much progress has been made in the last decade concerning
scattering and breakup reaction theory, core 
degrees of freedom  in the continuum has not been studied. 
The significance of the structure dynamics on reaction observables
can be very large (see for instance \cite{alkha}) and it is
fundamental to address this problem as soon as possible.
Given that there are large core excited configurations in many
dripline nuclei, one can expect an impact on many 
of the reaction observables. Work on the dynamic treatment of core excitation
in the continuum is underway \cite{summ05}.

\begin{theacknowledgments}
This work was supported by the National Superconducting Cyclotron Laboratory,
 Funda\c{c}\~ao para a Ci\^encia e a Tecnologia (F.C.T.)
of Portugal, under the grant POCTIC/36282/99, by the Department of Energy
under Grant No.\@ DE-FG03-93ER40773 and the U.\,S. National Science
Foundation under Grant No.\ PHY-0140343.
One of the authors (A.M.M.) acknowledges a F.C.T. post-doctoral grant.\end{theacknowledgments}

\bibliographystyle{aipprocl} 


\end{document}